%% file: lunghi.tex
\pdfoutput=1

\documentclass{PoS}
\usepackage{amsmath,amssymb,amsbsy,cite}
\usepackage{graphicx}
\usepackage{epsfig}
\usepackage{rotating} 

\def\tabvspace{\vphantom{$\Big($}}

\newcommand{\Delstar}{\ensuremath{\Delta^{\raise0.18ex\hbox{${\scriptstyle *}$}}}}
\def\gtwid{{\,\raise.35ex\hbox{$>$\kern-.75em\lower1ex\hbox{$\sim$}}\,}}
\def\ltwid{{\,\raise.35ex\hbox{$<$\kern-.75em\lower1ex\hbox{$\sim$}}\,}}
\def\leftvec{{\raise1.5ex\hbox{$\leftarrow$}\kern-1.00em}}
\def\rightvec{{\raise1.5ex\hbox{$\rightarrow$}\kern-1.00em}}
\def\half{{\scriptstyle \raise.2ex\hbox{${1\over2}$}}}
\def\threehalves{{\scriptstyle \raise.15ex\hbox{${3\over2}$}}}
\def\third{{\scriptstyle \raise.15ex\hbox{${1\over3}$}}}
\def\third{{\scriptstyle \raise.15ex\hbox{${1\over3}$}}}
\def\twothirds{{\scriptstyle \raise.15ex\hbox{${2\over3}$}}}
\def\fourth{{\scriptstyle \raise.15ex\hbox{${1\over4}$}}}

\newcommand*{\bea}{\begin{eqnarray}}
\newcommand*{\eea}{\end{eqnarray}}
\newcommand*{\be}{\begin{equation}}
\newcommand*{\ee}{\end{equation}}

\def\mev{{\rm MeV}}

\newcommand*{\CPT}{\raise0.4ex\hbox{$\chi$}PT}
\newcommand*{\chpt}{\raise0.4ex\hbox{$\chi$}PT}
\newcommand*{\schpt}{S\raise0.4ex\hbox{$\chi$}PT}

\def\Standard Modelb{\sigma_M}
\def\Standard Modelt{\overline{\sigma}_M}

\def\eqref#1{{(\ref{#1})}}

\def\bar{\overline}
\def\hat{\widehat}
\def\tilde{\widetilde}

\def\bea{\begin{eqnarray}}
\def\eea{\end{eqnarray}}

\def\beq{\begin{equation}}
\def\eeq{\end{equation}}

\def\spose#1{\hbox to 0pt{#1\hss}}
\def\ltapprox{\mathrel{\spose{\lower 3pt\hbox{$\mathchar"218$}}
 \raise 2.0pt\hbox{$\mathchar"13C$}}}
\def\gtapprox{\mathrel{\spose{\lower 3pt\hbox{$\mathchar"218$}}
 \raise 2.0pt\hbox{$\mathchar"13E$}}}
\def\inapprox{\mathrel{\spose{\lower 3pt\hbox{$\mathchar"218$}}
 \raise 2.0pt\hbox{$\mathchar"232$}}}

\title{Lessons for new physics from CKM studies }

\ShortTitle{Lessons for new physics from CKM studies}

\author{Jack Laiho\\
        Department of Physics and Astronomy, University of Glasgow, Glasgow, G128 QQ, UK\\
        E-mail: \email{jlaiho@fnal.gov}}

\author{\speaker{Enrico Lunghi}\\
       Indiana University, Bloomington, IN 47405 \\
       E-mail: \email{elunghi@indiana.edu}}

\author{Ruth S. Van de Water\\
        Physics Department, Brookhaven National Laboratory, Upton, NY 11973\\
        E-mail: \email{ruthv@bnl.gov}}

\abstract{We perform a global fit to the CKM unitarity triangle using the latest experimental and theoretical constraints.  We present results for three different sets of constraints:  the standard inputs used by CKMfitter~\cite{Charles:2004jd} and UTFit~\cite{Bona:2005vz}, the standard inputs minus $|V_{ub}|$, and the standard inputs minus both $|V_{ub}|$ and $|V_{cb}|$~\cite{Lunghi:2009ke}.  For the required nonperturbative weak matrix elements, we use three-flavor lattice QCD averages from \texttt {www.latticeaverages.org}; these have been updated from Ref.~\cite{Laiho:2009eu} to reflect all available lattice calculations as of the ``{\it End of 2010}".  Given current theoretical and experimental inputs, we observe an approximately 3$\sigma$ tension in the CKM unitarity triangle that can be interpreted as sign of physics beyond the standard model in the flavor sector.  Using a model-independent parameterization of new physics effects, we test the compatibility of new physics in kaon mixing, in $B$-mixing, or in $B\to\tau\nu$ decay with the current data. Although the tension could be accommodated with each hypothesis, the scenarios with new physics in $B$-mixing or, to a lesser extent, in $B\to\tau\nu$ decay are strongly preferred. Finally, we interpret these results in terms contributions to $\Delta S = 2$ and $\Delta B = 2$ four-fermion operators. We find that the preferred scale of new physics (with Standard Model like couplings) is in the few hundred GeV range.}

\FullConference{Flavor Physics and CP Violation 2010 \\
		 May 25-29,2010\\
		 Turin, Italy}

\begin{document}
\section{Motivation}

The B-factories and the Tevatron have produced a remarkable wealth of data needed to determine elements of the Cabibbo-Kobayashi-Maskawa (CKM) matrix and to search for new physics beyond the Standard Model CKM framework.  Despite the great experimental success of the Standard Model, there is now considerable evidence for physics beyond the Standard Model, such as dark matter, dark energy, and neutrino masses.  Generic new physics models to explain such phenomena also lead to additional $CP$-violating phases beyond the single one in the Standard Model;  this would lead to apparent inconsistencies between independent determinations of the CKM matrix elements.  Although there is presently reasonably good agreement with the Standard Model prediction of a single $CP$-violating phase, as measured by global fits of the CKM unitarity triangle, some tensions have been observed~\cite{Lunghi:2007ak, Lunghi:2008aa, Buras:2008nn, Buras:2009pj, Lenz:2010gu,Bona:2009cj,Lunghi:2009sm}.  In this work we use the latest theoretical and experimental inputs to quantify the tension with the Standard Model via a  global fit to the CKM unitarity triangle and then identify within a largely model-independent framework the most likely sources of the new physics.

\section{Unitarity Triangle Fit Preliminaries}
\label{sec:Prelim}

\subsection{Inputs}
\label{sec:inputs}
The standard analysis of the unitarity triangle involves a simultaneous fit to several quantities:  $\varepsilon_K$, $\Delta M_{B_d}$, $\Delta M_{B_s}$, time--dependent CP asymmetry in $B\to J/\psi K_s$ ($S_{\psi K} = \sin (2\beta)$, where $\beta$ is the phase of $V_{td}^*$),\footnote{A discussion of penguin pollution in $S_{\psi K}$ can be found in Ref.~\cite{Lunghi:2010gv}; see also Refs.~\cite{Boos:2004xp, Li:2006vq, Bander:1979px, Gronau:2008cc, Ciuchini:2005mg, Faller:2008zc, Ciuchini:2011kd}.} direct CP asymmetries in $B\to D^{(*)} K^{(*)}$ ($\gamma$ is the phase of $V_{ub}^*$) time dependent CP asymmetries in $B\to (\pi\pi, \rho\rho, \rho\pi)$ ($\alpha = \pi-\beta-\gamma$), ${\rm BR} (B\to\tau\nu)$,  $|V_{ub}|$ and $|V_{cb}|$ (from both inclusive and exclusive $b\to (u,c)\ell\nu$ with $\ell = e,\nu$).   We summarize the relevant inputs required for this analysis in Table~\ref{tab:inputs}. 

For the nonperturbative weak matrix elements we use averages of three-flavor lattice QCD calculations from Ref.~\cite{Laiho:2009eu} updated to reflect all results documented in proceedings or publications as of the ``{\it End of 2010}''.  We do not include two-flavor lattice calculations in our averages because of the unknown systematic error due to neglecting dynamical strange quark effects.  We treat both statistical and systematic errors as following a Gaussian distribution.  In performing the averages, we account for correlations between different lattice calculations of the same quantity in a reasonable but conservative manner;  whenever there is reason to believe that an error is correlated between two results, we assume that the degree of correlation is 100\%.  We obtain errors in pion and kaon matrix elements that are consistent with those of the Flavianet Lattice Averaging Group when we use the same inputs~\cite{Colangelo:2010et}, despite the fact that they use a different method for combining systematic errors between lattice calculations.  FLAG, however, has not yet updated their averages to reflect recent results from Lattice 2010 nor do they present averages of $B$- or $D$-meson quantities.  

There are several choices for how to implement the constraints from $B\to\tau\nu$ leptonic decay and $B_{d,s}$ mixing ($\Delta M_{B_d}$ and $\Delta M_{B_ds}$) on the unitarity triangle because one can parameterize the nonperturbative weak matrix element contributions to these quantities in different ways.  Certain combinations of lattice inputs are preferable, however, because they minimize correlations between the three different unitarity triangle constraints so that they can safely be neglected in the global fit. Let us now summarize the main considerations that lead to a reasonable choice of uncorrelated inputs. First of all it is important to include only one input with mass dimension 1 in order to eliminate correlations due to the determination of the lattice scale. Another important consideration is that the largest source of uncertainty in the $SU(3)$--breaking ratios, $\xi$ and $f_{B_s}/f_{B_d}$, is the chiral extrapolation.  Because the chiral logarithms are larger when the quark masses are lighter, the chiral extrapolation in the $SU(3)$-breaking ratios is primarily due to the chiral extrapolation in the $B_d$ quantities (i.e. $f_{B_d}$ and $\hat B_d$).  These ratios are, therefore, more correlated with $B_d$ rather than with $B_s$ quantities.  Finally we note that the decay constants $f_{B_d}$ and $f_{B_s}$ have a stronger chiral extrapolation than the $B$-parameters $\hat B_d$ and $\hat B_s$. In view of these considerations we choose to describe $B_q$ mixing in terms of $f_{B_s}\sqrt{\hat B_s}$ and $\xi$; once this choice is made, the additional input required to describe $B\to\tau\nu$ has to be $\hat B_d$ (because $f_{B_d}$ has mass dimension and is somewhat correlated with $\xi$). For completeness we point out that there is an alternative choice of inputs ($f_{B_s}/f_{B_d}$, $\hat B_s/\hat B_d$, $f_{B_s}$ and $\hat B_s$) for which correlations are again fairly small.

In our analysis, we write the unitarity triangle constraints in terms of the lattice inputs $f_{B_s}\sqrt{\hat B_s}$, $\xi$, and $\hat B_d$ such that the the unitarity triangle constraints are 
\begin{equation}
\Delta M_{B_d} \propto \left(\frac{ f_{B_s} \sqrt{\hat B_s}}{\xi}\right)^2 \; , \quad \quad
\Delta M_{B_s} \propto \left(f_{B_s} \textstyle \sqrt{\hat B_s}\right)^2 \; ,\quad \quad
{\rm BR} (B\to \tau \nu) \propto \frac{\left( f_{B_s} \sqrt{\hat B_s}\right)^2}{\xi^2 \hat B_d} \; .
\end{equation}
An additional advantage to this choice of inputs is that it allows us to use all existing experimental data to obtain a prediction for $f_{B_d}$ from all other constraints on the unitarity triangle that can be compared to the direct lattice calculation.

The determinations of $|V_{cb}|$ and $|V_{ub}|$ from inclusive and exclusive decays are problematic because they differ at the 2.1 and 1.7 $\sigma$ levels, respectively (note that if we remove the additional $10\%$ model uncertainty on inclusive $|V_{ub}|$, the discrepancy rises to the 3.3 $\sigma$ level).

The determinations of $|V_{cb}|$ and $|V_{ub}|$ from inclusive and exclusive decays are problematic because they differ at the 2.1 and 1.7 s levels, respectively (note that if we remove the additional $10\%$ model uncertainty on inclusive $|V_{ub}|$, the discrepancy rises to the $3.3 \sigma$ level).  For this reason, when we combine the inclusive and exclusive determinations of $|V_{cb}|$ and $|V_{ub}|$ we inflate the errors by the square root of the chi-square per degree-of-freedom (as prescribed by the PDG);  the resulting averages are given in the bottom panel of Table~\ref{tab:inputs}.  Furthermore, in addition to the standard fit in which all measurements are included, we consider two additional scenarios in which we remove $|V_{ub}|$ and $|V_{cb}|$ from the chi-square.  The strategy for removing $|V_{cb}|$ by combining the constraints from $\varepsilon_K$, $\Delta M_{B_s}$, and ${\rm BR} (B\to \tau \nu)$ was proposed and is described in detail in Ref.~\cite{Lunghi:2009ke}.

\begin{table}[t]
\begin{center}
\setlength{\unitlength}{1cm}
\begin{picture}(1, 1)
\put(0.4, 1.6){ \begin{sideways} Lattice QCD inputs \end{sideways}}
\put(0.4, -2.98){ \begin{sideways} Other inputs \end{sideways}}
\end{picture}
\begin{tabular}{ll}
\hline
\hline
$\left| V_{cb} \right|_{\rm excl} =(39.5 \pm 1.0) \times 10^{-3}$&
$\left| V_{ub} \right|_{\rm excl} = (3.12 \pm 0.26) \times 10^{-3} $ \tabvspace \\
$\hat B_K = 0.737 \pm 0.020$ &
$\kappa_\varepsilon = 0.94 \pm 0.02$ \tabvspace \\
$f_B = (205 \pm 12) \; \mev$ &
$f_{B_s} = (250 \pm 12) \; \mev$  \tabvspace \\
$\hat B_{B_d} = 1.26 \pm 0.11$ &
$\hat B_{B_s} = 1.33 \pm 0.06$ \tabvspace \\
$f_{B_d} \sqrt{\hat B_{B_d}} = (233 \pm 14) \; \mev $ & 
$f_{B_s} \sqrt{\hat B_{B_s}} = (288 \pm 15) \; \mev $ \tabvspace \\
$\xi  \equiv f_{B_s}\sqrt{\hat B_s}/(f_{B_d}\sqrt{\hat B_d}) = 1.237 \pm 0.032$ & 
$f_{B_s}/f_{B_d} = 1.215 \pm 0.019$\tabvspace\\ 
\hline\hline
$\left| V_{cb} \right|_{\rm incl} =(41.68 \pm 0.44 \pm 0.09 \pm 0.58) \times 10^{-3}$~\cite{HFAG_FPCP_09}&
$\alpha = (89.5 \pm 4.3)^{\rm o}$ \tabvspace\\
$\left| V_{ub} \right|_{\rm incl} = (4.34 \pm 0.16^{+0.15}_ {-0.22}\pm 0.43) \times 10^{-3} $~\cite{HFAG_FPCP_09}  &
 $\eta_1 = 1.51 \pm 0.24$~\cite{Herrlich:1993yv}  \vphantom{\Big(} \\
${\rm BR} (B\to \tau\nu) = (1.68 \pm 0.31) \times 10^{-4}$~\cite{Ikado:2006un,Sanchez:2010rt,Hara:2010dk} & 
$S_{\psi K_S} = 0.668 \pm 0.023$~\cite{Kreps:2010ts} \\
$\Delta m_{B_d} = (0.507 \pm 0.005)\; {\rm ps}^{-1}$~\cite{HFAG_PDG_09} & 
$\gamma = (78 \pm 12)^{\rm o}$~\cite{Bona:2005vz,Bona:2006ah} \vphantom{\Big(} \\
$\Delta m_{B_s} = (17.77 \pm 0.10 \pm 0.07)\;  {\rm ps}^{-1}$~\cite{Evans:2007hq} & 
 $\eta_2 = 0.5765 \pm 0.0065$~\cite{Buras:1990fn}  \vphantom{\Big(}\\
$m_{t, pole} = (172.4 \pm 1.2) \; {\rm GeV}$~\cite{:2008vn} & 
$\eta_3 = 0.494 \pm 0.046$~\cite{Herrlich:1995hh,Brod:2010mj}  \vphantom{\Big(}\\
$m_c(m_c) = (1.268 \pm 0.009 ) \; {\rm GeV}$~\cite{Allison:2008xk}&  
$\eta_B = 0.551 \pm 0.007$~\cite{Buchalla:1996ys} \vphantom{\Big(} \\
$\varepsilon_K = (2.229 \pm 0.012 ) \times 10^{-3}$~\cite{Yao:2006px} &
$\lambda = 0.2253  \pm 0.0009$~\cite{Antonelli:2010yf}\vphantom{\Big(}  \\ 
\hline
\hline
$\left| V_{cb} \right|_{\rm avg} =(40.77 \pm 0.81) \times 10^{-3}$ &
$\left| V_{ub} \right|_{\rm avg} = (3.37 \pm 0.49) \times 10^{-3}$ \tabvspace \\
\hline
\hline
\end{tabular}
\caption{Lattice QCD and other inputs to the unitarity triangle analysis. The determination of $\alpha$ is obtained from a combined isospin analysis of $B\to (\pi\pi,\; \rho\rho, \; \rho\pi)$ branching ratios and CP asymmetries~\cite{HFAG_PDG_09}. References for the lattice-QCD results entering the averages in the the upper panel can be found in Ref.~\cite{Laiho:2009eu} with updates at \texttt{www.latticeaverages.org}.  Updated lattice averages for other quantities that do not enter the global unitarity triangle fit such as pion, kaon, and $D$-meson decay constants and light-quark masses can also be found at \texttt{www.latticeaverages.org}. \label{tab:inputs}}
\end{center}
\end{table}

\subsection{Interpretation as New Physics}

We interpret the observed tensions in the global unitarity triangle fit as contributions from new physics in either in kaon mixing, $B_d$-mixing, or $B\to \tau \nu$;  details of the analysis method are given in Ref.~\cite{Lunghi:2009sm}.  We adopt a model-independent parametrization of new physics effects in the three observables:
\bea
|\varepsilon_K^{\rm NP}| &=& C_\varepsilon \; |\varepsilon_K^{\rm SM}| \; , \\
M_{12}^{d,{\rm NP}} &=& r_d^2 \; e^{2 i \theta_d} \; M_{12}^{d,{\rm SM}} \;, \\
{\rm BR} (B\to \tau\nu)^{\rm NP} &=&  \left(
1- \frac{\tan^2 \beta \; m_{B^+}^2}{m_{H^+}^2 (1+\epsilon_0 \tan\beta)} \right) {\rm BR} (B\to \tau\nu)^{\rm SM} \; \\
&=&r_H \;  {\rm BR} (B\to\tau\nu)^{\rm SM}  \; ,
\eea
where in the Standard Model ($C_\varepsilon, \; r_H, \; r_d )= 1$ and $\theta_d = 0$. In presence of non-vanishing contributions to $B_d$ mixing the following other observables are also affected:
\bea
S_{\psi K_S} &=& \sin 2 (\beta + \theta_d)  \; , \\
\sin (2 \alpha_{\rm eff})  &=& \sin 2 (\alpha - \theta_d)  \; , \\
X_{sd} &=& \frac{\Delta M_{B_s}}{\Delta M_{B_d}} = X_{sd}^{\rm SM} \; r_d^{-2} \; .
\eea
When considering new physics in $B_d$ mixing we allow simultaneous variations of both $\theta_d$ and $r_d$. We find that new physics in $|M_{12}^d|$ has a limited effect on the tension between the direct and indirect determinations of $\sin (2\beta)$; as a consequence, our results for $r_d$ and $\theta_d$ point to larger effects on the latter. 

Finally we interpret the constraints on the parameters $C_\varepsilon, \; r_d$, and $\theta_d$ in terms of generic new physics contributions to $\Delta S = 2$ and $\Delta B = 2$ four-fermion operators. The most general effective Hamiltonian for $B_d$--mixing can be written as~\footnote{The Hamiltonians for $B_s$-- and $K$--mixing are obtained by replacing $(d,b) \to (s,b)$ and $(d,b) \to (d,s)$, respectively.}
\bea
{\cal H}_{\rm eff} = \frac{G_F^2 m_W^2}{16 \pi^2} \left( V_{tb}^{} V_{td}^*\right)^2 \left( 
\sum_{i=1}^5 C_i  O_i  +  \sum_{i=1}^3 \tilde C_i  \tilde O_i \right) 
\eea
where
\begin{eqnarray}
\begin{tabular}{lcl}
$O_1 =  ( \bar d_L \gamma_\mu b_L) ( \bar d_L \gamma_\mu b_L)$
& \phantom{ciaciacia} &
$\tilde O_1  =  ( \bar d_R \gamma_\mu b_R) ( \bar d_R \gamma_\mu b_R)$
\cr
$O_2  =  ( \bar d_R  b_L) ( \bar d_R  b_L) $
& &
$\tilde O_2  =  ( \bar d_L  b_R) ( \bar d_L  b_R)$
\cr
$O_3  =  ( \bar d^{\alpha}_R  b_L^\beta ) ( \bar d^{\beta}_R  b_L^\alpha )$
& &
$\tilde O_3  =  ( \bar d^{\alpha}_L b_R^\beta ) ( \bar d^{\beta}_L  b_R^\alpha ) $
\cr
$O_4  =  ( \bar d_R  b_L) ( \bar d_L  b_R) $ 
& &
$O_5  =  ( \bar d^{\alpha}_R  b_L^\beta ) ( \bar d^{\beta}_L  b_R^\alpha )$ .
\cr
\end{tabular}
\end{eqnarray}
Within the Standard Model, only the operator $O_1$ receives a non-vanishing contribution at a high scale $\mu_H \sim m_t$. For our analysis we assume that all new physics effects can be effectively taken into account by a suitable contribution to $C_1$:
\bea
{\cal H}_{\rm eff} = \frac{G_F^2 m_W^4}{16 \pi^2} \left( V_{tb}^{} V_{td}^*\right)^2 C_1^{\rm SM} 
\left( \frac{1}{m_W^2} - \frac{e^{i\varphi}}{ \Lambda^2} \right) O_1 \; ,
\label{np}
\eea
where the minus sign has been introduced {\it a posteriori} (as we will see the the fit will point to new physics phases of order $\varphi \sim O(1)$). In this parametrization $\Lambda$ is the scale of some new physics model whose interactions are identical to the Standard Model with the exception of an additional arbitrary CP violating phase:  
\bea
C_1 = C_1^{\rm SM} \left( 1 - e^{i\varphi} \frac{m_W^2}{\Lambda^2} \right) \;.
\label{lamphi}
\eea
When discussing new physics in the kaon sector, we will consider also a similar new physics contribution to the operator $O_4$: because of RG effects and of the chiral enhancement of the matrix element of $O_4$, the latter usually point to a new physics scale that is larger by a factor $\sim 65$ than for $O_1$ case.

\section{Unitarity Triangle Fit Results and Constraints on New Physics}
In this section we present the results we obtain for the full fit and for the fits in which semileptonic decays (for the extraction of $|V_{ub|}$ and $V_{cb}|$) are not used.  For each set of constraints we present the fitted values of the CKM parameters $\bar\rho$, $\bar\eta$ and $A$. We also show the predictions for several interesting quantities (most importantly $S_{\psi K}$ and ${\rm BR} (B\to \tau\nu)$) that we obtain after removing the corresponding direct determination from the fit.  Finally we interpret the observed discrepancies in terms of new physics in $\varepsilon_K$, $B_d$--mixing or $B\to \tau\nu$. 
\begin{figure}[t]
\begin{center}
\includegraphics[width=0.6 \linewidth]{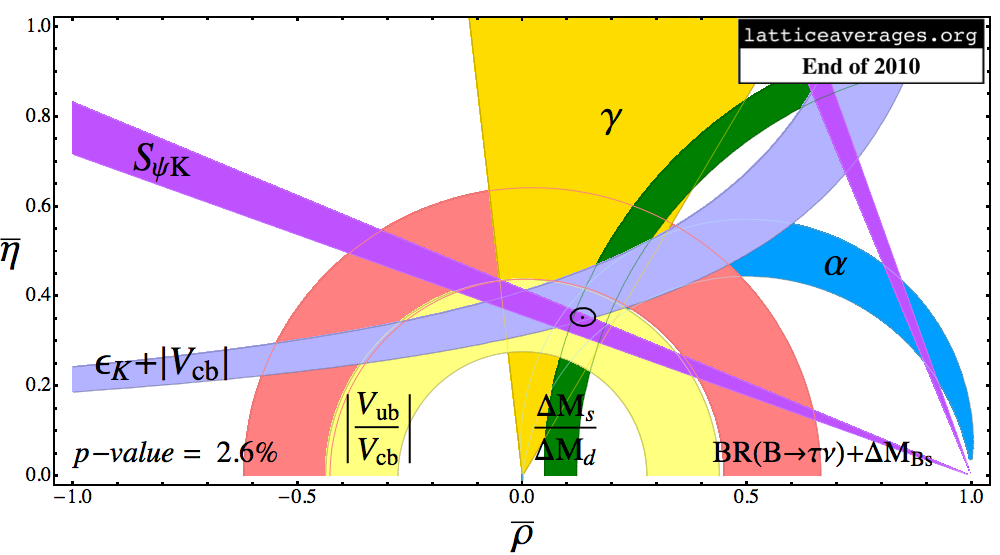}
\includegraphics[width=0.6 \linewidth]{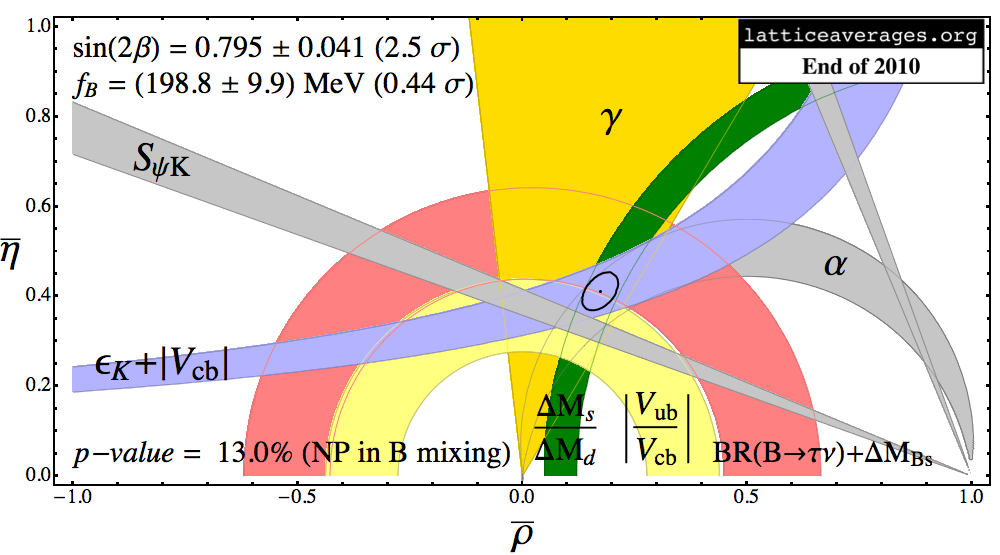}
\includegraphics[width=0.6\linewidth]{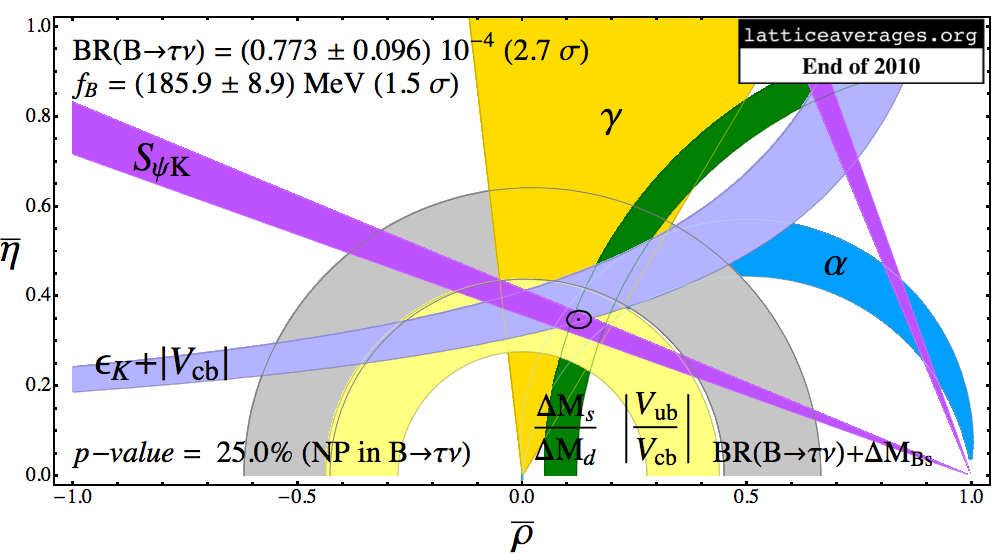} 
\caption{Unitarity triangle fit with all constraints included. Quantities that are not used to generate the black contour are grayed out.\label{fig:utfit-tot}}
\end{center}
\end{figure}

\begin{figure}[t]
\begin{center}
\includegraphics[width=0.35 \linewidth]{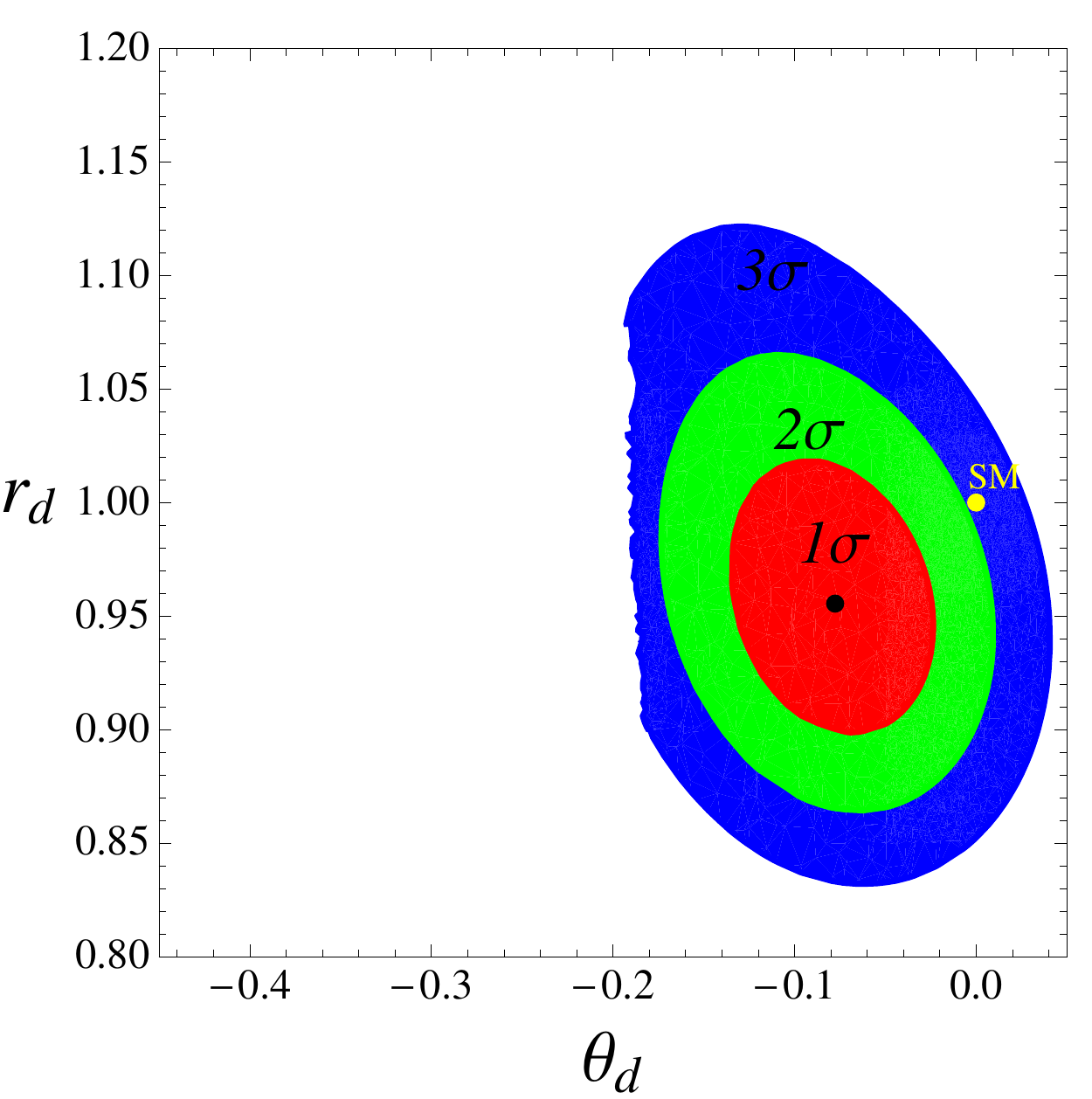}
\includegraphics[width=0.35 \linewidth]{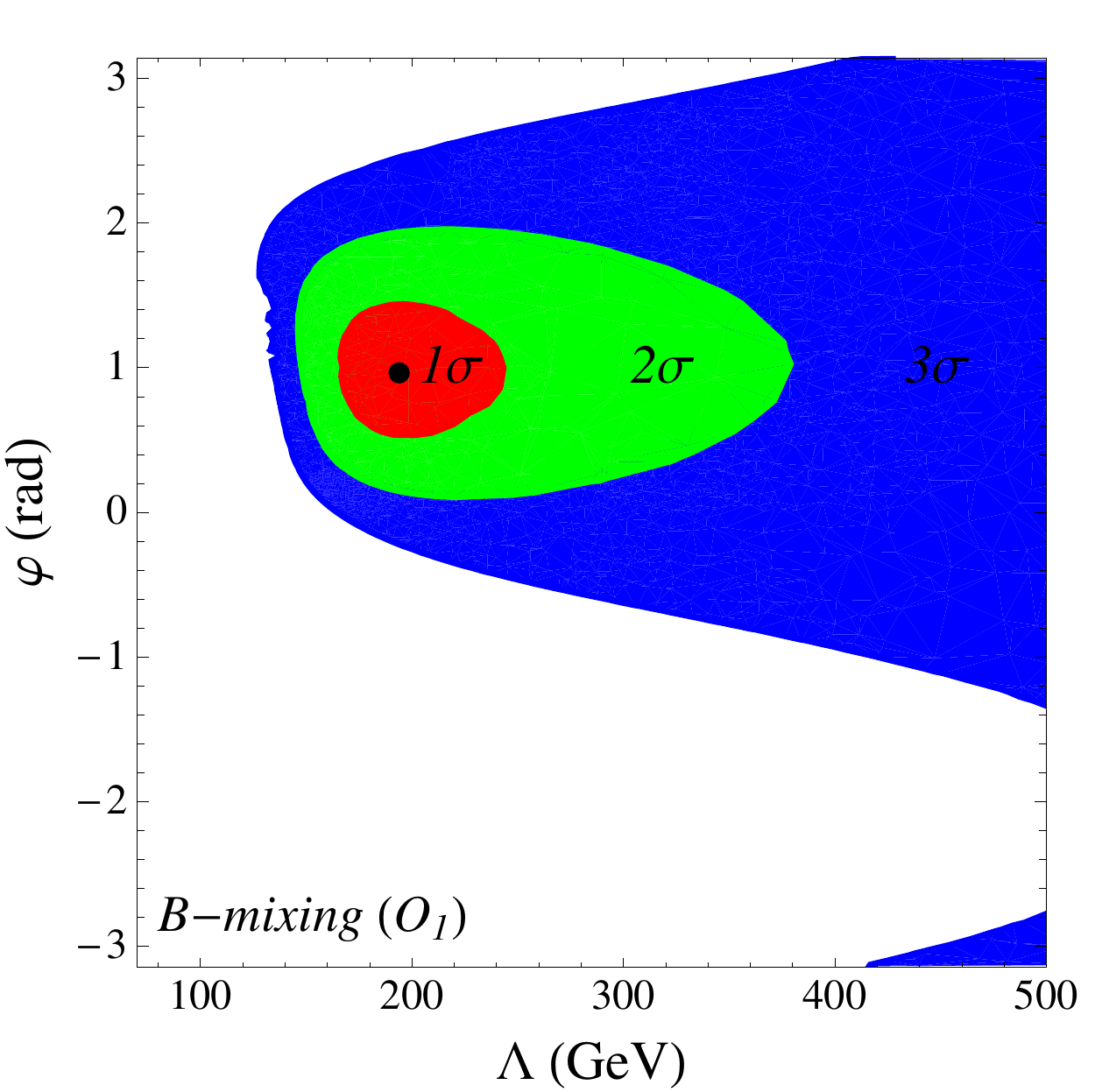}
\includegraphics[width=0.35 \linewidth]{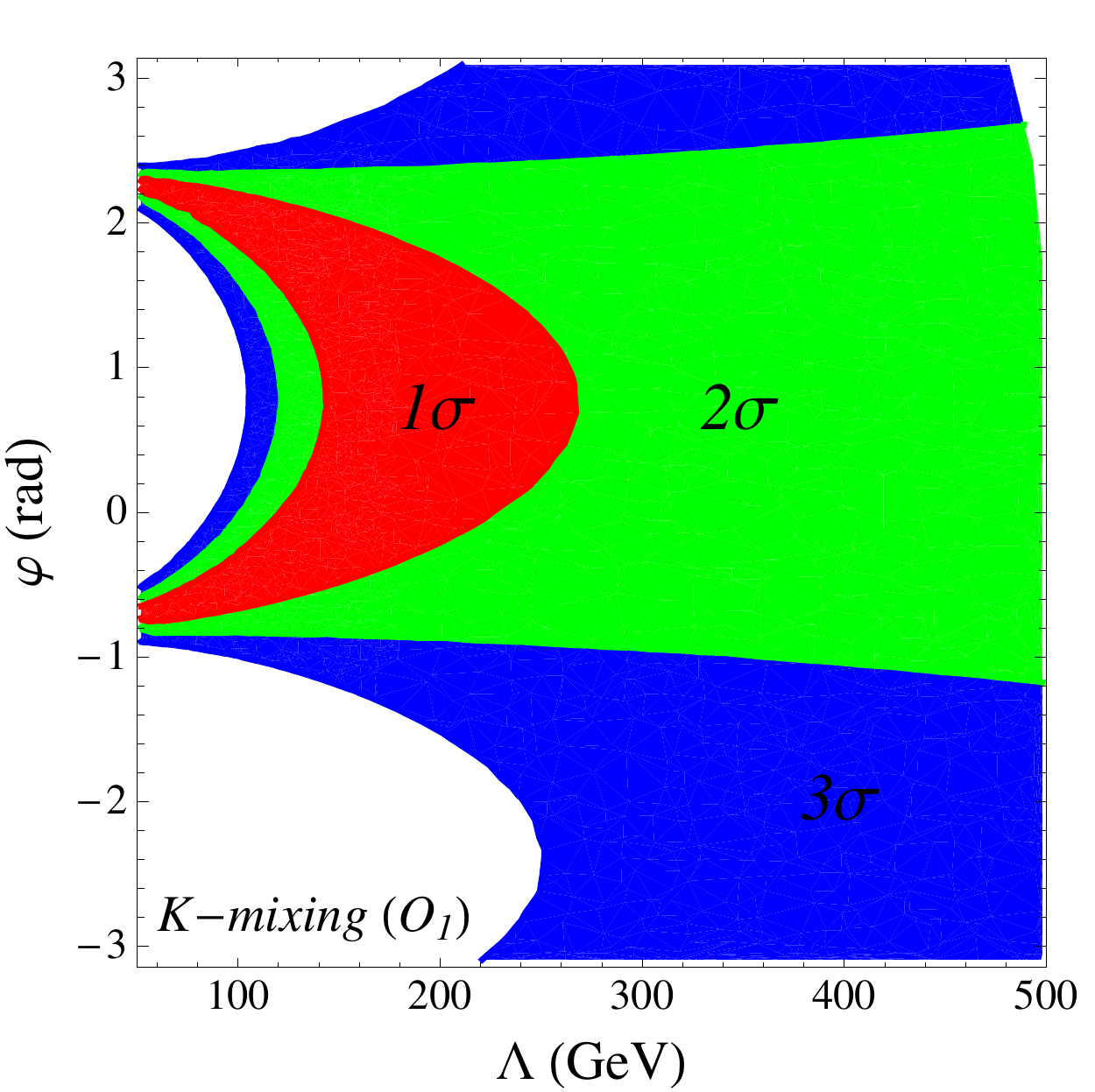}
\includegraphics[width=0.35 \linewidth]{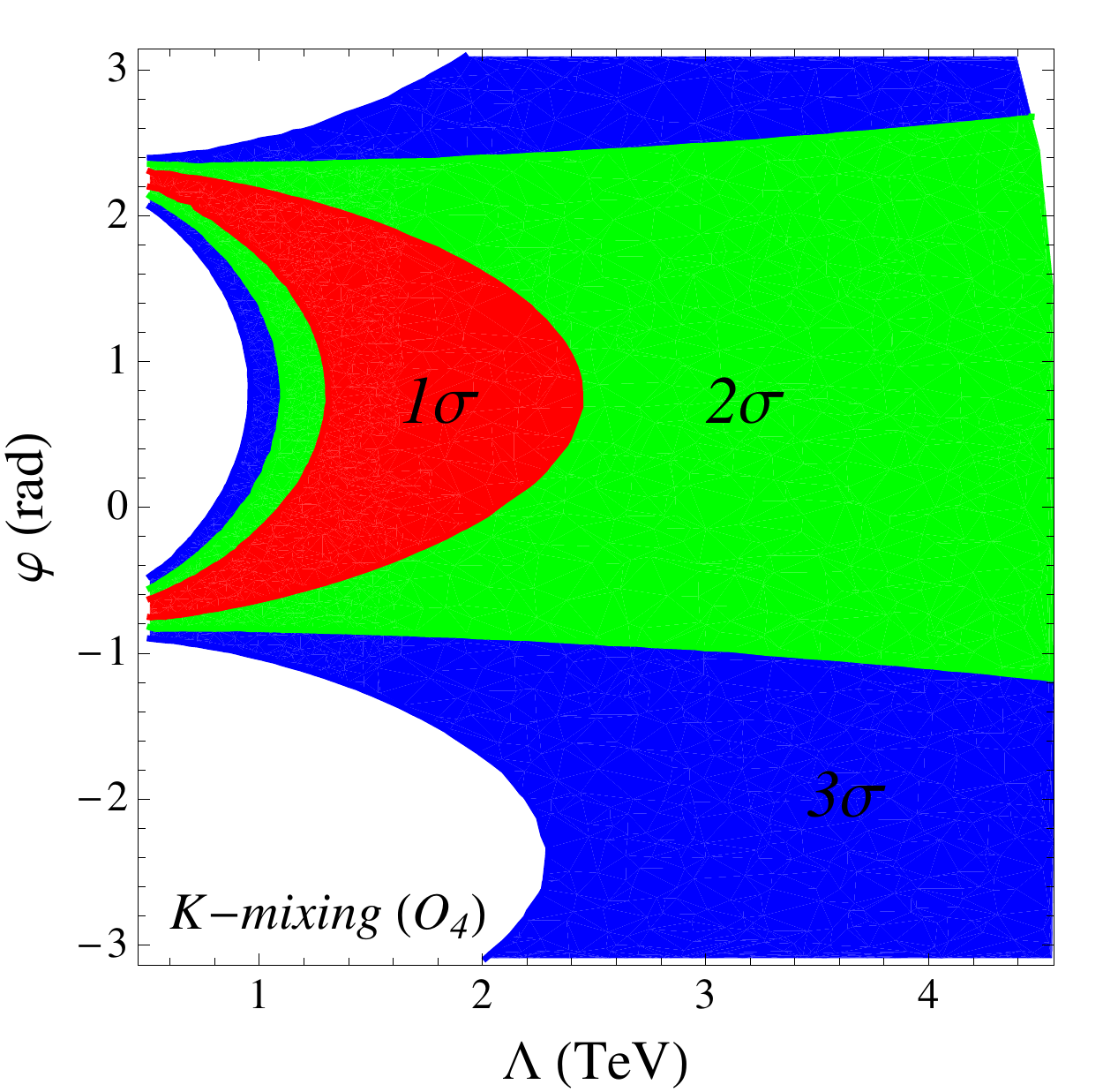}
\caption{Unitarity triangle fit with all constraints included: new physics analysis. \label{fig:utfit-tot-NP}}
\end{center}
\end{figure}

\begin{figure}[t]
\begin{center}
\includegraphics[width= 0.6 \linewidth]{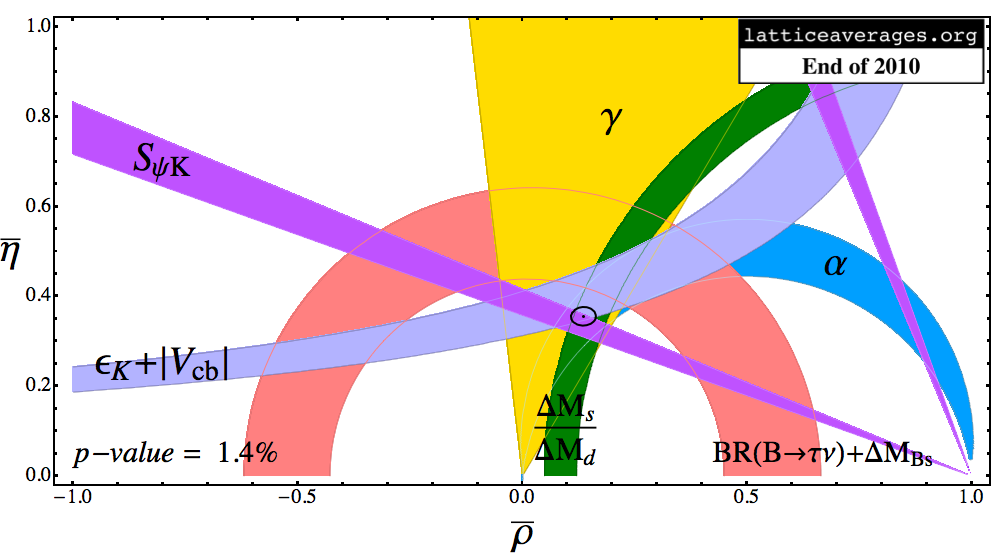}
\includegraphics[width= 0.6 \linewidth]{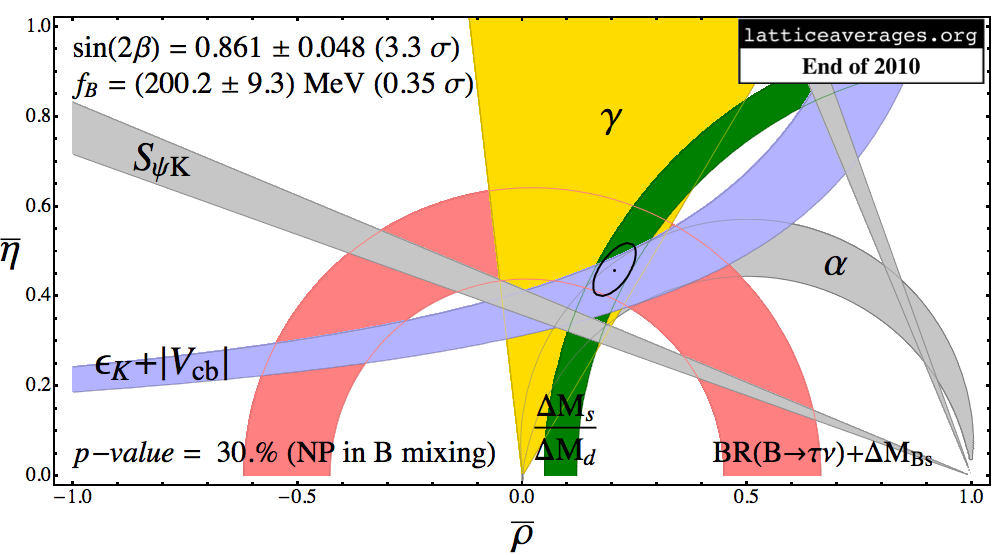}
\includegraphics[width= 0.6 \linewidth]{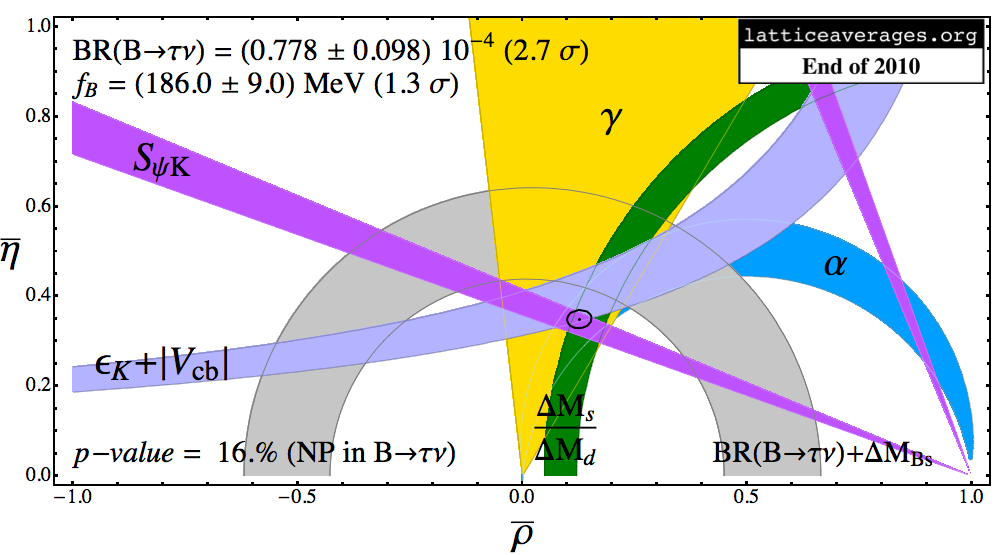}
\caption{Unitarity triangle fit without $|V_{ub}|$. Quantities that are not used to generate the black contour are grayed out.\label{fig:utfit-novub}}
\end{center}
\end{figure}

\begin{figure}[t]
\begin{center}
\includegraphics[width=0.35 \linewidth]{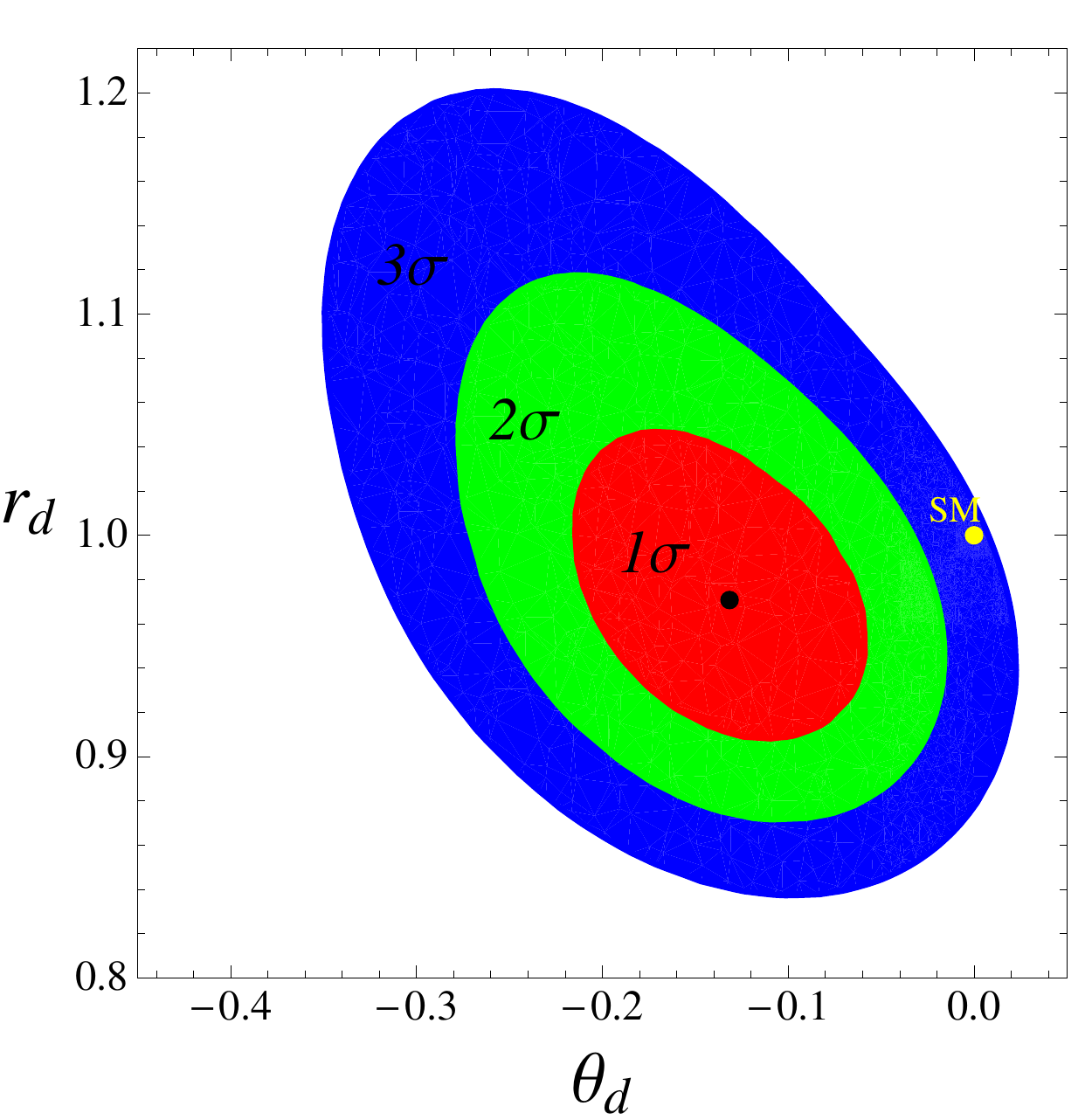}
\includegraphics[width=0.35 \linewidth]{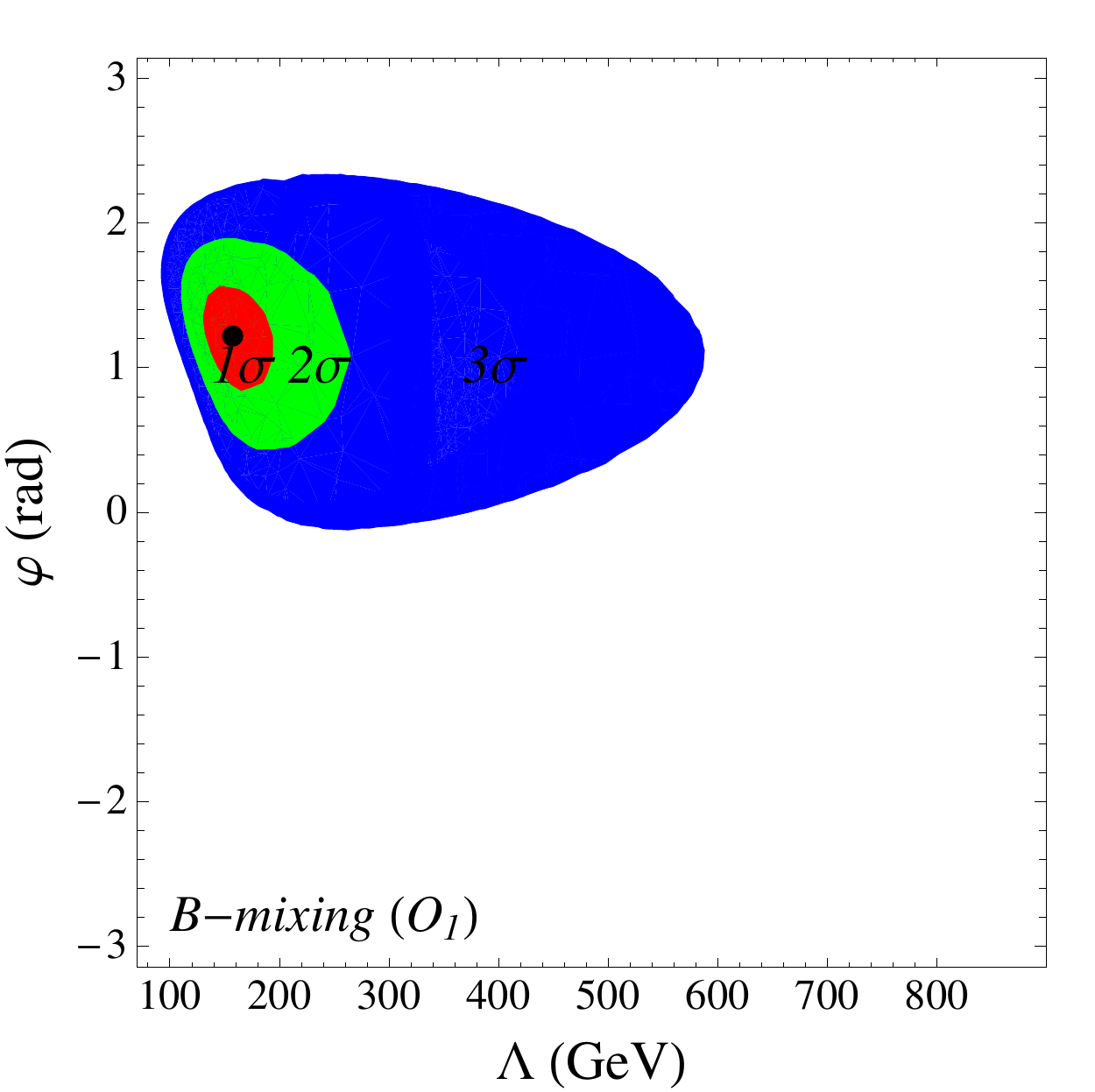}
\includegraphics[width=0.35 \linewidth]{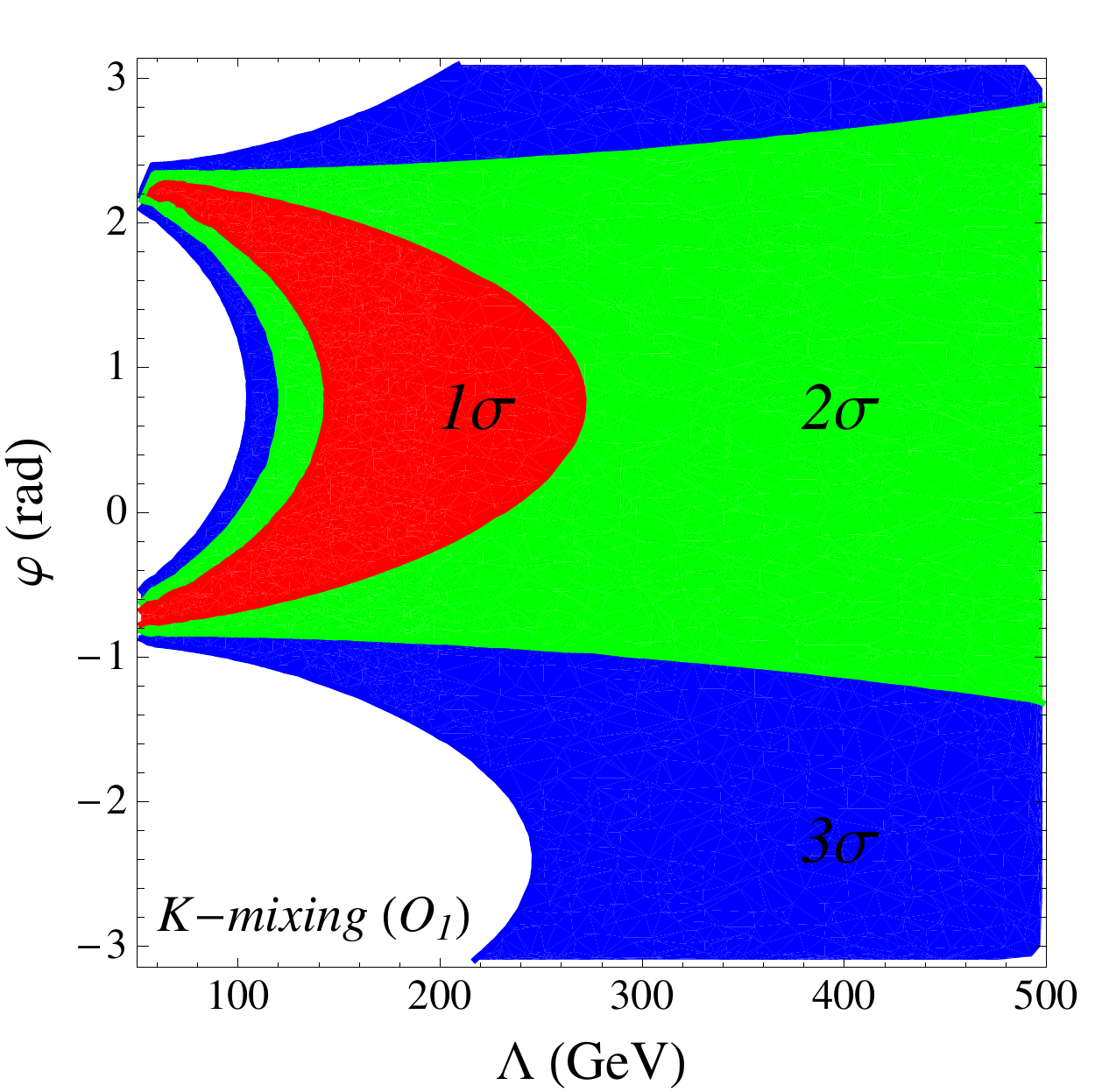}
\includegraphics[width=0.35 \linewidth]{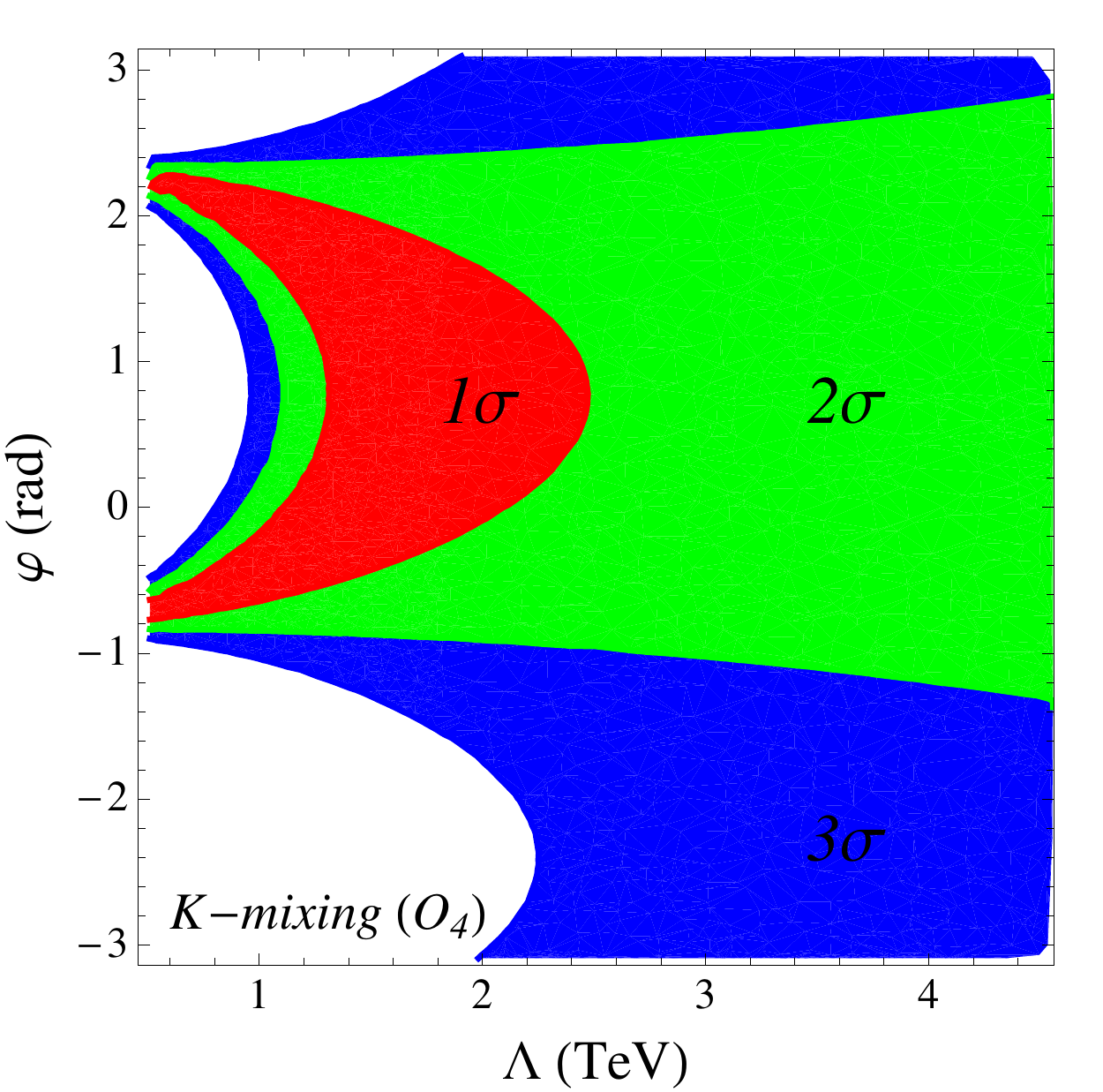}
\caption{Unitarity triangle fit  without $|V_{ub}|$: new physics analysis. \label{fig:utfit-novub-NP}}
\end{center}
\end{figure}

\begin{figure}[t]
\begin{center}
\includegraphics[width=0.6 \linewidth]{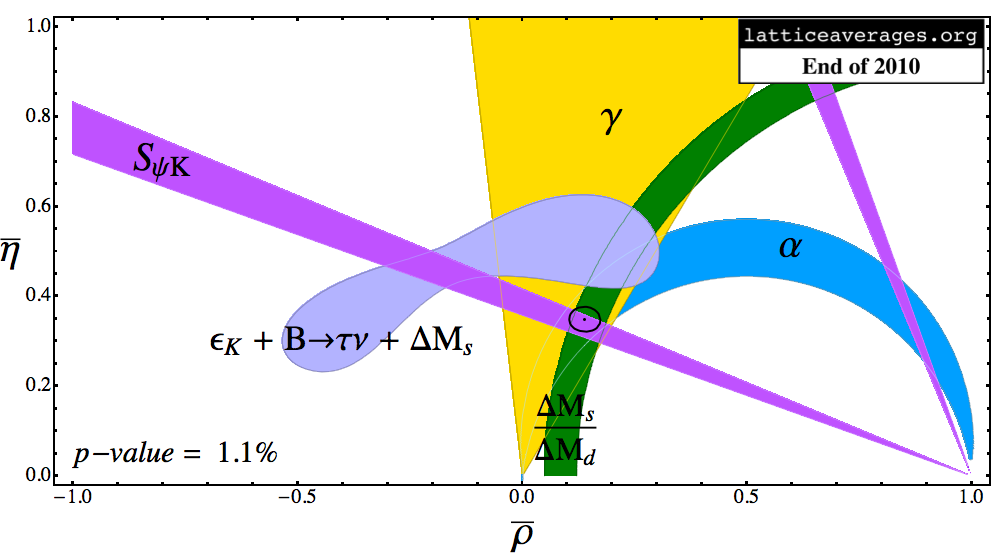} 
\includegraphics[width=0.6 \linewidth]{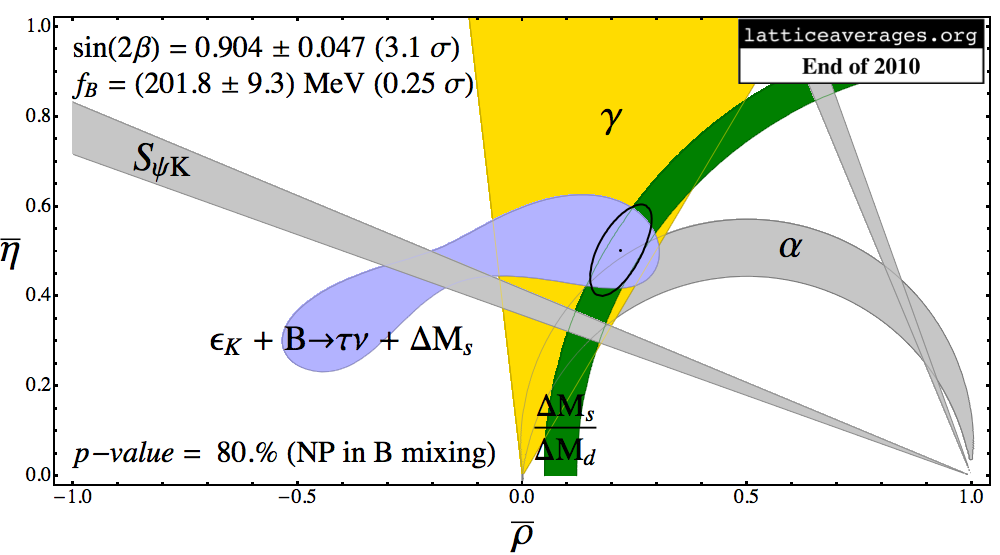} 
\includegraphics[width=0.6 \linewidth]{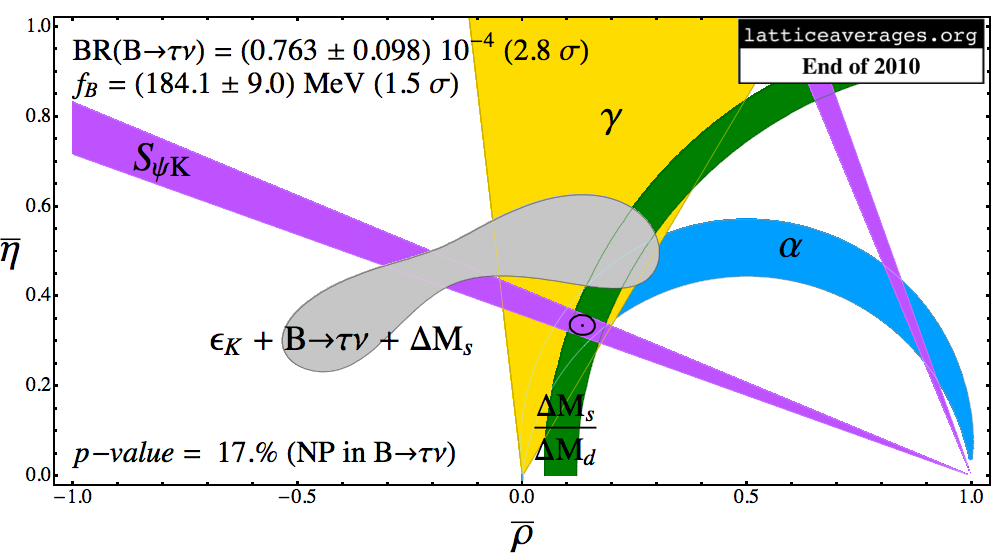} 
\caption{Unitarity triangle fit without $|V_{ub}|$ and $|V_{cb}|$. Quantities that are not used to generate the black contour are grayed out.\label{fig:utfit-novqb}}
\end{center}
\end{figure}

In Figs.~\ref{fig:utfit-tot}, \ref{fig:utfit-novub} and \ref{fig:utfit-novqb}, we show the global CKM unitarity triangle fit for the three set of inputs that we consider (complete fit, no $V_{ub}$ fit, no $V_{qb}$ fit). In each figure, the black contours and $p$--values in the top, middle and bottom panels correspond to the complete fit, the fit with a new phase in $B$ mixing (i.e. without using $S_{\psi K}$ and $\alpha$) and the fit with new physics in $B\to \tau \nu$ (i.e. without using ${\rm BR} (B\to\tau\nu)$), respectively.\footnote{Note that in the no $V_{qb}$ fit we define the new physics in $B\to\tau\nu$ scenario by removing $B\to\tau\nu$, $\Delta M_{B_s}$ and $\varepsilon_K$ from the fit.} In the fits with a new phase in $B$ mixing we also show the fit predictions for $\sin(2\beta)$ and $f_B$; in the fits with new physics in $B\to \tau \nu$, we show the fit predictions for ${\rm BR} (B\to\tau\nu)$ and $f_B$. Note that the individual contours in Figs.~\ref{fig:utfit-tot}-\ref{fig:utfit-novqb} never use the same input twice in order to minimize the correlations between constraints: in particular, the $B\to\tau\nu$ allowed area is obtained by using $\Delta M_{B_s}$ instead of the direct determination of $|V_{cb}|$.

\begin{figure}[t]
\begin{center}
\includegraphics[width=0.35 \linewidth]{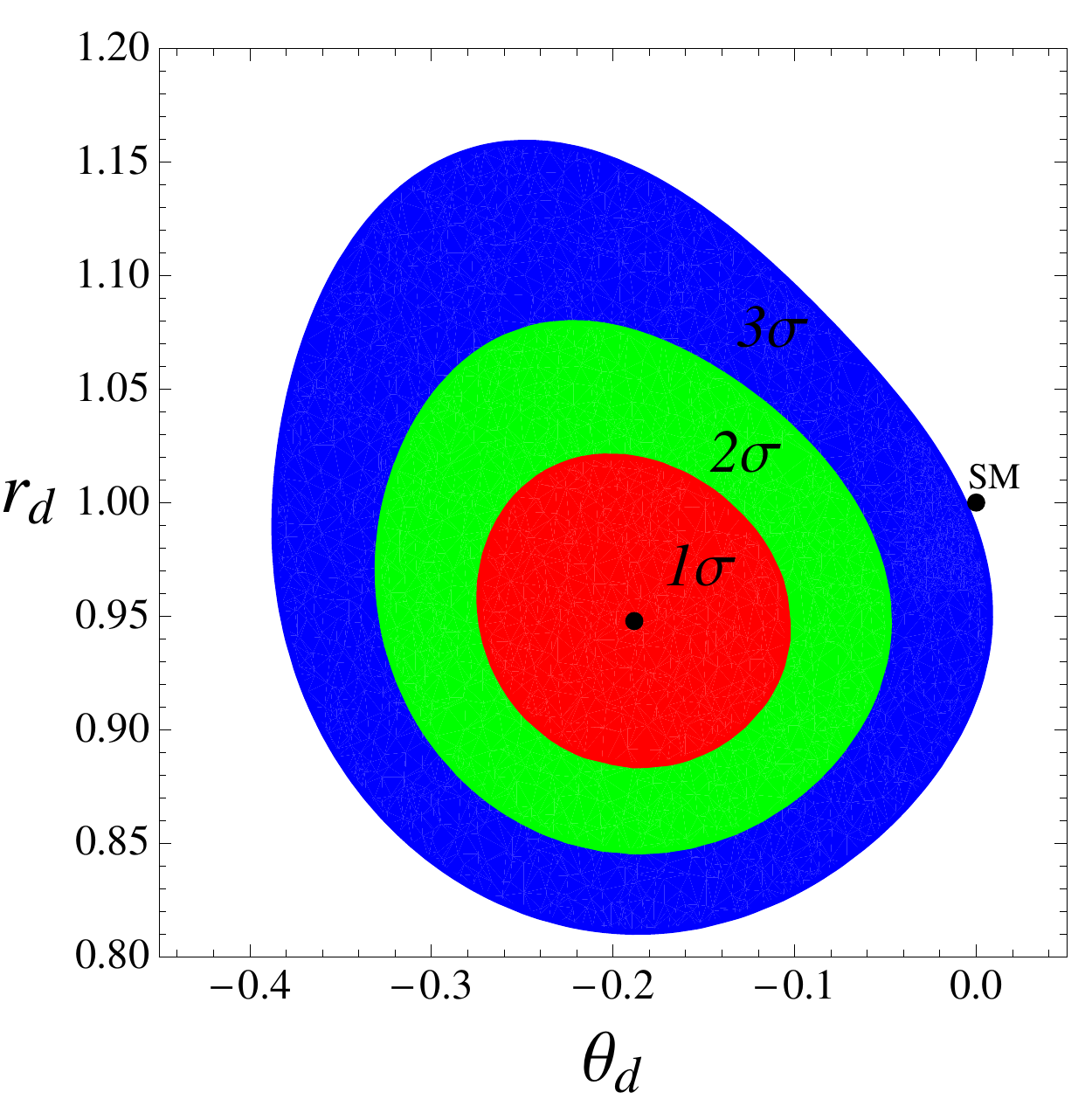}
\includegraphics[width=0.35 \linewidth]{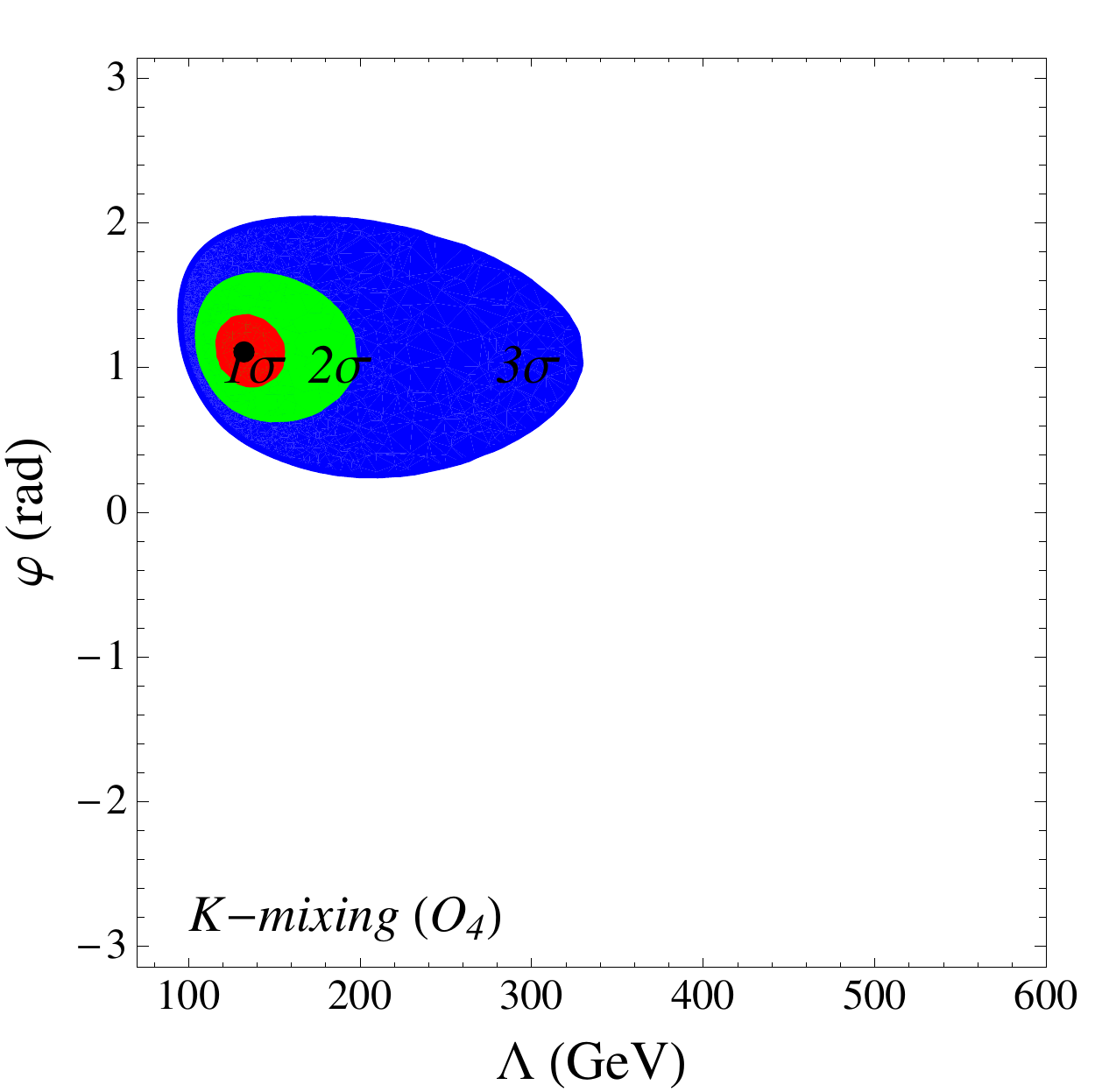}
\includegraphics[width=0.35 \linewidth]{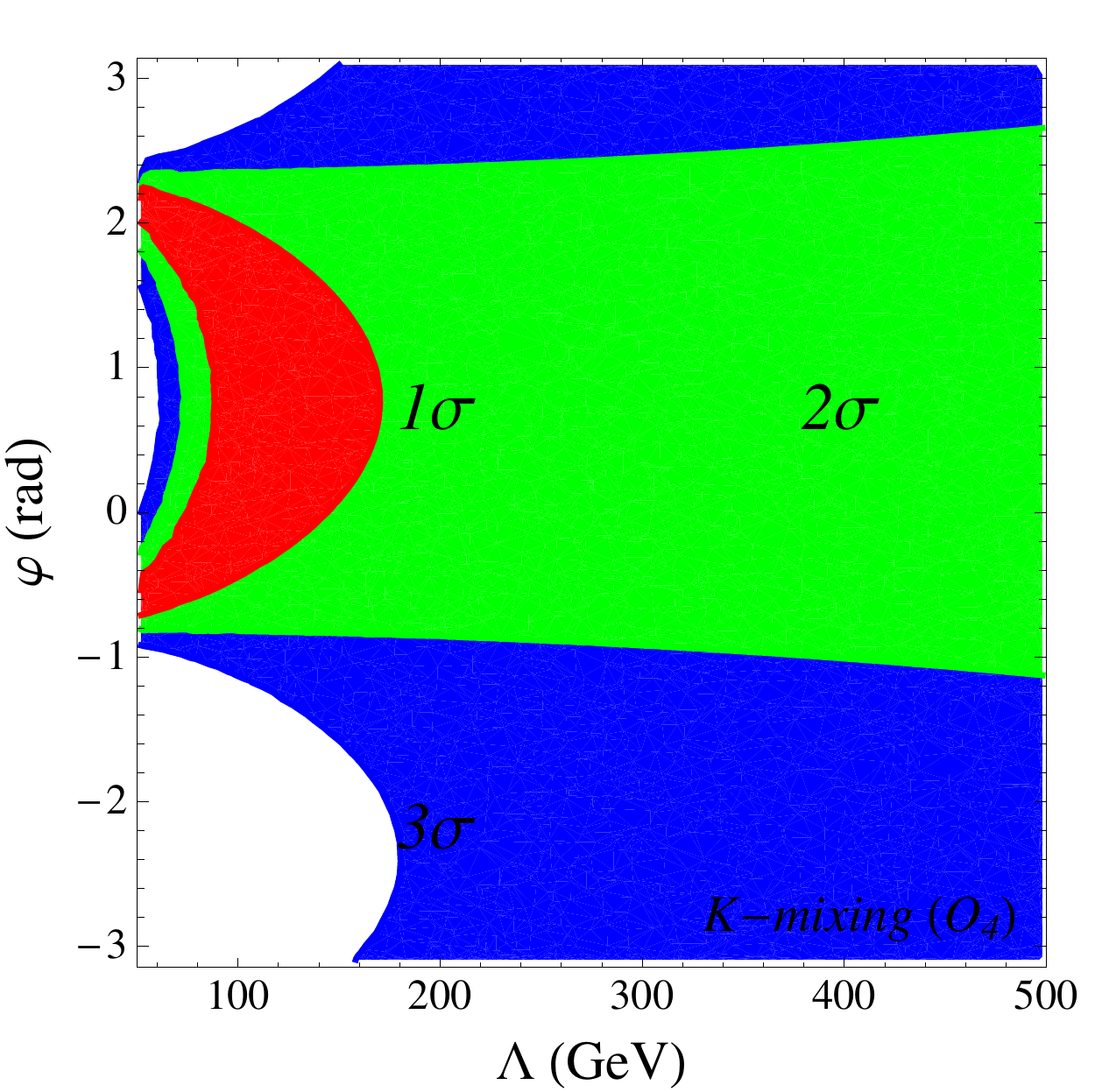}
\includegraphics[width=0.35 \linewidth]{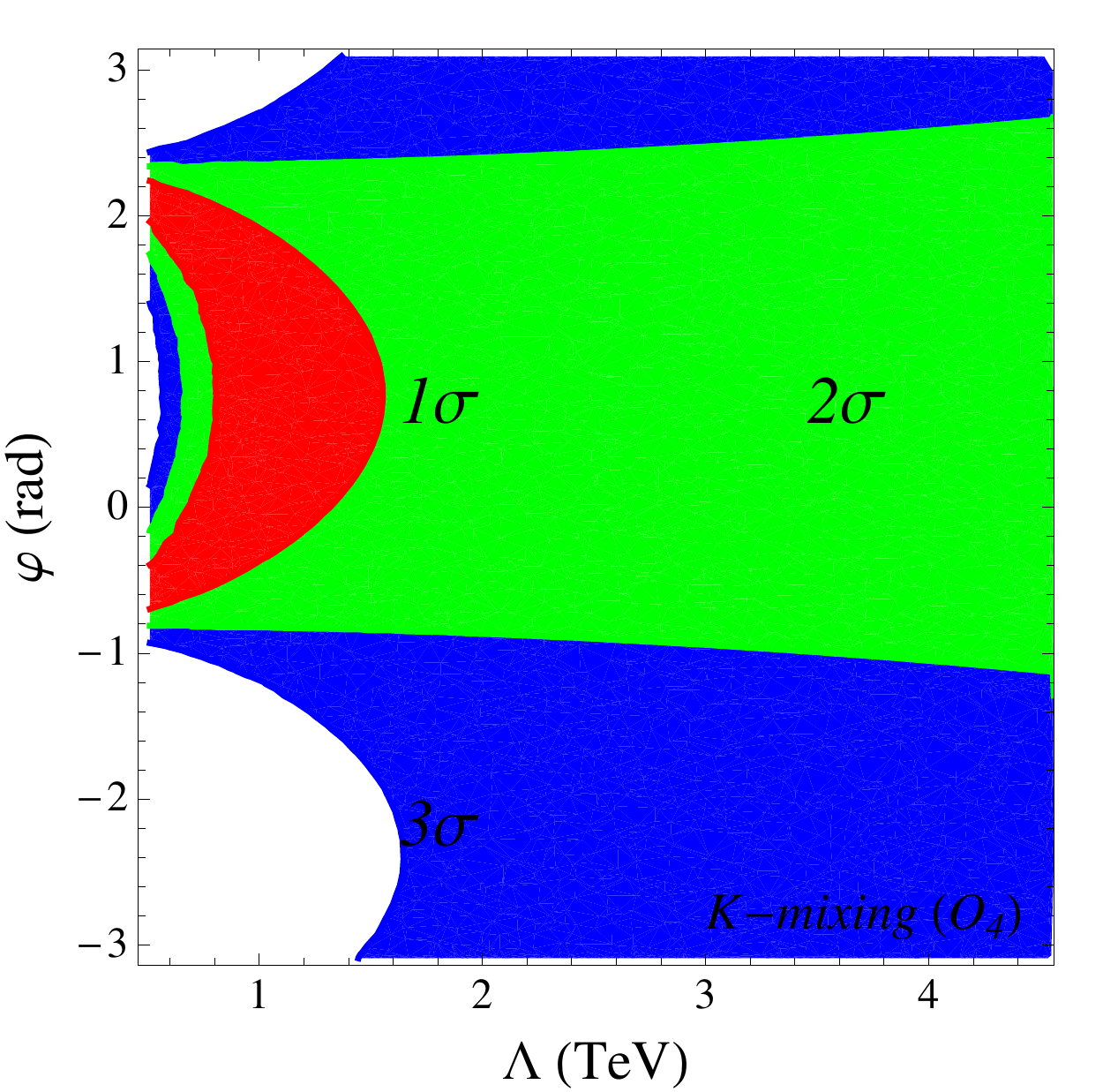}
\caption{Unitarity triangle fit without $|V_{ub}|$ and $|V_{cb}|$: new physics analysis.\label{fig:utfit-novqb-NP}}
\end{center}
\end{figure}

In Figs.~\ref{fig:utfit-tot-NP}, \ref{fig:utfit-novub-NP} and \ref{fig:utfit-novqb-NP} we show the interpretation of the tensions highlighted in the various fits in terms of possible new physics. In the first panel of each figure we show the result of the two-dimensional fit in the $(\theta_d,r_d)$ plane and in the second panel we map this allowed region onto the $(\Lambda,\varphi)$ plane (see Eq.~(\ref{lamphi})) under the assumption of new physics in $O_1$ only. In the third and fourth panels we show the allowed $(\Lambda,\varphi)$ regions for the scenario with new physics in $K$ mixing (the left and right panels show contributions to $O_1$ and $O_4$, respectively).
\subsection{Standard Fit}
\noindent We include constraints from $\varepsilon_K$, $\Delta M_{B_d}$, $\Delta M_{B_s}$, $\alpha$, $S_{\psi K}$, $\gamma$, ${\rm BR} (B\to\tau\nu)$, $|V_{cb}|$ and $|V_{ub}|$. The overall $p$-value of the Standard Model fit is $ \input pvalue-complete.txt $ and the results of the fit are
\beq
\input RHO-complete.txt \quad\quad 
\input ETA-complete.txt \quad\quad 
\input A-complete.txt  
\; .
\eeq
The predictions from all other information when the direct determination of the quantity is removed from fit are
\begin{align}
& \input VUB-complete.txt \\
& \input APSI-complete.txt \\
& \input VCB-complete.txt \\
& \input BK-complete.txt \\
& \input FBSSQRTBD-complete.txt \\
& \input BTN-complete.txt \\
&
\hskip -0.5cm
\begin{cases}
\input FB-complete.txt  & \text{complete fit} \\
\input FBNOS2B-complete.txt & \text{without using} \; S_{\psi K}  \\
\input FBNOBTN-complete.txt &\text{without using} \; {\rm BR}(B\to\tau\nu) \\
\end{cases}
\label{fbtot}
\end{align}
where we indicate the deviation from the corresponding direct determination in parentheses.
The interpretation of the above discrepancies in terms of new physics in $K$--mixing, $B_d$--mixing and $B\to \tau \nu$ yields
\begin{align}
& \hphantom{5\;\;} \input CEPSILON-complete.txt   \\ 
&\begin{cases}
\input THETAD-complete.txt   \\
\input RD-complete.txt   \\ 
\end{cases}  ( \input THETAD1-complete.txt )\\
&\hphantom{5\;\;}  \input RH-complete.txt 
\; .
\end{align}
Figure~\ref{fig:utfit-tot} summarizes these results. 
\subsection{Fit without $|V_{ub}|$}
\label{sec:fitnovub}
\noindent We include constraints from $\varepsilon_K$, $\Delta M_{B_d}$, $\Delta M_{B_s}$, $\alpha$, $S_{\psi K}$, $\gamma$, ${\rm BR} (B\to\tau\nu)$ and $|V_{cb}|$. The overall $p$-value of the Standard Model fit is $ \input pvalue-novub.txt $ and the results of the fit are
\beq
\input RHO-novub.txt \quad\quad 
\input ETA-novub.txt \quad\quad 
\input A-novub.txt  
\; .
\eeq
The predictions from all other information when the direct determination of the quantity is removed from fit are
\begin{align}
& \input VUB-novub.txt \\
& \input APSI-novub.txt \\
& \input VCB-novub.txt \\
& \input BK-novub.txt \\
& \input FBSSQRTBD-novub.txt \\
& \input BTN-novub.txt \\
&
\hskip -0.5cm
\begin{cases}
\input FB-novub.txt  & \text{no $V_{ub}$ fit} \\
\input FBNOS2B-novub.txt & \text{without using} \; S_{\psi K}  \\
\input FBNOBTN-novub.txt &\text{without using} \; {\rm BR}(B\to\tau\nu) \\
\end{cases}
\label{fbnovub}
\end{align}
where we indicate the deviation from the corresponding direct determination in parentheses.
The interpretation of the above discrepancies in terms of new physics in $K$--mixing, $B_d$--mixing and $B\to \tau \nu$ yields
\begin{align}
& \hphantom{5\;\;} \input CEPSILON-novub.txt   \label{cepsilon} \\ 
&\begin{cases}
\input THETAD-novub.txt   \\
\input RD-novub.txt   \\ 
\end{cases}  ( \input THETAD1-novub.txt )\\
&\hphantom{5\;\;}  \input RH-novub.txt \label{rh}
\; .
\end{align}
Figure~\ref{fig:utfit-novub} summarizes these results. 
\subsection{Fit without $|V_{ub}|$ and $|V_{cb}|$}
\noindent We include constraints from $\varepsilon_K$, $\Delta M_{B_d}$, $\Delta M_{B_s}$, $\alpha$, $S_{\psi K}$, $\gamma$ and ${\rm BR} (B\to\tau\nu)$. The overall $p$-value of the Standard Model fit is $  \input pvalue-novqb.txt  $ and the results of the fit are
\beq
\input RHO-novqb.txt \quad\quad 
\input ETA-novqb.txt \quad\quad 
\input A-novqb.txt  
\; .
\eeq
The predictions from all other information when the direct determination of the quantity is removed from fit are
\begin{align}
& \input VUB-novqb.txt \\
& \input APSI-novqb.txt \\
& \input VCB-novqb.txt \\
& \input BK-novqb.txt \\
& \input FBSSQRTBD-novqb.txt \\
& \input BTN-novqb.txt \\
&
\hskip -0.5cm
\begin{cases}
\input FB-novqb.txt  & \text{no $V_{qb}$ fit} \\
\input FBNOS2B-novqb.txt & \text{without using} \; S_{\psi K}  \\
\input FBNOBTN-novqb.txt &\text{without using} \; {\rm BR}(B\to\tau\nu) \\
\end{cases}
\label{fbnovqb}
\end{align}
where we indicate the deviation from the corresponding direct determination in parentheses.
The interpretation of the above discrepancies in terms of new physics in $K$--mixing, $B_d$--mixing and $B\to \tau \nu$ yields
\begin{align}
& \hphantom{5\;\;} \input CEPSILON-novqb.txt   \\ 
&\begin{cases}
\input THETAD-novqb.txt   \\
\input RD-novqb.txt   \\ 
\end{cases}  ( \input THETAD1-novqb.txt )\\
&\hphantom{5\;\;}  \input RH-novqb.txt 
\; .
\end{align}
Figure~\ref{fig:utfit-novqb} summarizes these results. 

\section{Discussion}
The main lessons to be discerned from the fits described in the previous sections are:
\begin{itemize}
\item If we take the current experimental and theoretical errors in the inputs to the global CKM unitarity triangle fit at face value, then the CKM description of flavor and $CP$ violation displays a tension at the 3$\sigma$ level.
\item The use of $b\to u \ell \nu \; (\ell = e,\mu)$ semileptonic decays is problematic. The $(2- 3)\sigma$ discrepancy between the inclusive and exclusive extractions of $|V_{ub}|$ when coupled to the complexity of theoretical methods used is cause of serious concern. Furthermore, the direct determination of ${\rm BR} (B\to \tau \nu)$ from $f_B$ and $|V_{ub}|_{\rm excl}$ (we obtain $(0.64 \pm 0.12) \times 10^{-4}$) is $3.1\sigma$ below its direct experimental determination (see Table~\ref{tab:inputs}). These two $3\sigma$ tensions in the $B\to \tau\nu$, $f_B$,  $|V_{ub}|$ system could be quite naturally resolved by a shift in the direct extraction of $|V_{ub}|$ due to either new physics (see for instance Refs.~\cite{Crivellin:2009sd,Buras:2010pz}) or further theoretical progress on inclusive and exclusive semileptonic decays. In the light of these considerations we believe that ``fit without $|V_{ub}|$'' presented in Sec.~\ref{sec:fitnovub} is the most apt to represent our present understanding of $CP$ violation in the CKM. 
\item The fit seems to prefer an interpretation of this $3\sigma$ tension in terms of new physics in $B_d$ mixing or $B\to\tau\nu$ rather than in $\varepsilon_K$ (see the p-values in Eqs.~(\ref{cepsilon}-\ref{rh})). Note that even removing ${\rm BR} (B\to\tau\nu)$ from the fit without $|V_{ub}|$, we obtain $S_{\psi K}^{\rm fit}  =  0.803 \pm 0.069 \; (1.9 \sigma)$. Thus the presence of new physics in $B\to\tau\nu$ can reduce the deviation of $S_{\psi K}^{\rm fit}$ from $S_{\psi K}^{\rm exp}$ from $3.2 \sigma $ to $1.9 \sigma$, but cannot completely eliminate the tension in the global unitarity triangle fit. In contrast, new physics in $B_d$ mixing  can completely tackle the problem: the values of $\rho$ and $\eta$ required to eliminate the $1.9\sigma$ tension, also yield a ${\rm BR} (B\to\tau\nu)$ that is in perfect agreement with experiments. Therefore we conclude that the most satisfactory resolution of the overall tension is through new physics in $B_d$ mixing. This conclusion is reinforced by the inspection of Eqs.~(\ref{fbtot}), (\ref{fbnovub}) and (\ref{fbnovqb}): the fit result for $f_{B_d}$ is much closer to direct lattice determination in the fit with new physics in $B_d$ mixing (i.e. without using $S_{\psi K}$) rather than in $B\to\tau\nu$ (i.e. without using ${\rm BR} (B\to \tau\nu)$).\footnote{See Ref.~\cite{Lunghi:2010gv} for a comprehensive discussion of this point.}
\item In terms of a new physics model whose interactions mimic closely the SM (see Eq.~(\ref{np})), this tension points to a few hundred GeV mass scale. Even allowing for a generous model dependence in the couplings, it seems that such new particles, if the tension in the fit stands confirmed, cannot escape detection in direct production experiments.
\end{itemize}

\end{document}

%% file: pvalue-complete.txt
p = 2.6\%

%% file: RHO-complete.txt
\bar \rho  =  0.135 \pm 0.018

%% file: ETA-complete.txt
\bar \eta  =  0.354 \pm 0.013

%% file: A-complete.txt
A  =  0.816 \pm 0.011

%% file: VUB-complete.txt
|V_{ub}|  =  (3.64 \pm 0.13 ) \; \times 10^{-3} \quad (0.53\; \sigma)

%% file: APSI-complete.txt
S_{\psi K}  =  0.795 \pm 0.041 \quad (2.5\; \sigma)

%% file: VCB-complete.txt
|V_{cb}|  =  (42.1 \pm 0.82 ) \; \times 10^{-3} \quad (1.1\; \sigma)

%% file: BK-complete.txt
\hat B_K  =  0.889 \pm 0.083 \quad (1.9\; \sigma)

%% file: FBSSQRTBD-complete.txt
f_{B_d} \sqrt{\hat B_d}  = (210.0 \pm 4.3 ) \; {\rm MeV} 

%% file: BTN-complete.txt
{\rm BR} (B\to\tau\nu)  =  (0.773 \pm 0.096 ) \; \times 10^{-4} \quad (2.7\; \sigma)

%% file: FB-complete.txt
f_{B_d}  =  (193. \pm 10. ) \; {\rm MeV} \quad (0.8\; \sigma)

%% file: FBNOS2B-complete.txt
f_{B_d}  =  (198.8 \pm 9.9 ) \; {\rm MeV} \quad (0.4\; \sigma)

%% file: FBNOBTN-complete.txt
f_{B_d} = (185.9 \pm 8.9 ) \; {\rm MeV} \quad (1.\; \sigma)

%% file: CEPSILON-complete.txt
C_\varepsilon  =  1.21 \pm 0.12 \quad\quad ( 1.9\; \sigma ,\;  p = 0.056)

%% file: THETAD-complete.txt
\theta_d  =  - (4.5 \pm 2.1)^{\rm o}

%% file: RD-complete.txt
r_d  =  0.96 \pm 0.039

%% file: THETAD1-complete.txt
2.2\; \sigma ,\;  p = 0.13

%% file: RH-complete.txt
r_H  =  2.22 \pm 0.49 \quad\quad ( 2.8\; \sigma ,\;  p = 0.25)

%% file: pvalue-novub.txt
p = 1.4\%

%% file: RHO-novub.txt
\bar \rho  =  0.136 \pm 0.018

%% file: ETA-novub.txt
\bar \eta  =  0.355 \pm 0.013

%% file: A-novub.txt
A  =  0.816 \pm 0.011

%% file: VUB-novub.txt
|V_{ub}|  =  (3.64 \pm 0.13 ) \; \times 10^{-3} \quad (0.53\; \sigma)

%% file: APSI-novub.txt
S_{\psi K} = 0.861 \pm 0.048 \quad (3.3\; \sigma)

%% file: VCB-novub.txt
|V_{cb}| = (42.04 \pm 0.82 ) \; \times 10^{-3} \quad (1.1\; \sigma)

%% file: BK-novub.txt
\hat B_K = 0.887 \pm 0.083 \quad (1.9\; \sigma)

%% file: FBSSQRTBD-novub.txt
f_{B_d} \sqrt{\hat B_d}  =  (210.2 \pm 4.3 ) \; {\rm MeV} 

%% file: BTN-novub.txt
{\rm BR} (B\to\tau\nu)  =  (0.778 \pm 0.098 ) \; \times 10^{-4} \quad (2.7\; \sigma)

%% file: FB-novub.txt
f_{B_d} = (194. \pm 10. ) \; {\rm MeV} \quad (0.8\; \sigma)

%% file: FBNOS2B-novub.txt
f_{B_d} = (200.2 \pm 9.3 ) \; {\rm MeV} \quad (0.4\; \sigma)

%% file: FBNOBTN-novub.txt
f_{B_d} = (186.0 \pm 9.0 ) \; {\rm MeV} \quad (1.3\; \sigma)

%% file: CEPSILON-novub.txt
C_\varepsilon  =  1.20 \pm 0.12 \quad\quad ( 1.9\; \sigma ,\;  p = 0.030)

%% file: THETAD-novub.txt
\theta_d  =  - (7.7 \pm 3.0)^{\rm o}

%% file: RD-novub.txt
r_d  =  0.97 \pm 0.045

%% file: THETAD1-novub.txt
2.9\; \sigma ,\;  p = 0.30

%% file: RH-novub.txt
r_H  =  2.20 \pm 0.49 \quad\quad ( 2.8\; \sigma ,\;  p = 0.16)

%% file: pvalue-novqb.txt
p = 1.1\%

%% file: RHO-novqb.txt
\bar \rho =  0.139 \pm 0.018

%% file: ETA-novqb.txt
\bar \eta  =  0.349 \pm 0.015

%% file: A-novqb.txt
A  =  0.828 \pm 0.016

%% file: VUB-novqb.txt
|V_{ub}|  =  (3.64 \pm 0.13 ) \; \times 10^{-3} \quad (0.53\; \sigma)

%% file: APSI-novqb.txt
S_{\psi K}  =  0.904 \pm 0.047 \quad (3.1\; \sigma)

%% file: VCB-novqb.txt
|V_{cb}|  =  (42.1 \pm 0.82 ) \; \times 10^{-3} \quad (1.1\; \sigma)

%% file: BK-novqb.txt
\hat B_K  =  1.11 \pm 0.21 \quad (1.9\; \sigma)

%% file: FBSSQRTBD-novqb.txt
f_{B_d} \sqrt{\hat B_d}  =  (208.2 \pm 4.6 ) \; {\rm MeV} 

%% file: BTN-novqb.txt
{\rm BR} (B\to\tau\nu)=(0.763 \pm 0.098 ) \; \times 10^{-4} \quad (2.8\; \sigma)

%% file: FB-novqb.txt
f_{B_d}  =  (192. \pm 10. ) \; {\rm MeV} \quad (1.3\; \sigma)

%% file: FBNOS2B-novqb.txt
f_{B_d}  =  (201.8 \pm 9.3 ) \; {\rm MeV} \quad (0.2\; \sigma)

%% file: FBNOBTN-novqb.txt
f_{B_d}  =  (184.1 \pm 9.0 ) \; {\rm MeV} \quad (1.5\; \sigma)

%% file: CEPSILON-novqb.txt
C_\varepsilon =  1.51 \pm 0.29 \quad ( 2.0\; \sigma ,\;  p = 0.027)

%% file: THETAD-novqb.txt
\theta_d  = - (10.8 \pm 3.2)^{\rm o}

%% file: RD-novqb.txt
r_d  =  0.95 \pm 0.045

%% file: THETAD1-novqb.txt
3.3\; \sigma ,\;  p = 0.80

%% file: RH-novqb.txt
r_H  =  2.25 \pm 0.50 \quad ( 2.8\; \sigma ,\;  p = 0.17)